


\magnification=\magstep1

\raggedbottom

\rightline{\vbox{
\hbox{MPI-PhT/94-75}
\hbox{DAMTP R94/43}
\hbox{gr-qc/9411008}
\hbox{November 1994}}}
\bigskip\bigskip

\font\bigboldd=cmbx10 scaled \magstep3
\centerline{\bigboldd A Parallelizable Implicit Evolution Scheme}
\centerline{\bigboldd for Regge Calculus}
\bigskip
\hfuzz=33pt
\centerline{\bf
                John W.~Barrett\footnote{$^1$}{Mathematics Department, The
University of Nottingham, University Park, Nottingham NG7 2RD, UK},
                Mark Galassi\footnote{$^2$}{Space Data Systems Group, Los
Alamos National Laboratory, Los Alamos, NM 87545, USA},
                Warner A.~Miller\footnote{$^3$}{Theoretical Division, Los
Alamos National Laboratory, Los Alamos, NM 87545, USA},
}
\centerline{\bf
                Rafael D.~Sorkin\footnote{$^4$}{Physics Department,
Syracuse University, Syracuse, NY 13244--1130, USA},
                Philip A.~Tuckey\footnote{$^5$}{Max--Planck--Institut f\"ur
Physik, F\"ohringer Ring 6, D-80805 M\"unchen, Germany}
            and Ruth M.~Williams\footnote{$^6$}{DAMTP, Silver Street,
Cambridge CB3 9EW, UK}
}
\bigskip
\bigskip
\centerline{\bf ABSTRACT}

The role of Regge calculus as a tool for numerical relativity is discussed,
and a parallelizable implicit evolution scheme described.  Because of the
structure of the Regge equations, it is possible to advance the vertices of
a triangulated spacelike hypersurface in isolation, solving at each vertex
a purely local system of implicit equations for the new edge-lengths
involved.  (In particular, equations of global ``elliptic-type'' do not
arise.)  Consequently, there exists a parallel evolution scheme which
divides the vertices into families of non-adjacent elements and advances
all the vertices of a family simultaneously.  The relation between the
structure of the equations of motion and the Bianchi identities is also
considered.  The method is illustrated by a preliminary application to a
600--cell Friedmann cosmology.  The parallelizable evolution algorithm
described in this paper should enable Regge calculus to be a viable
discretization technique in numerical relativity.

\vfill
\eject

\noindent
\line{\bf 1. Numerical relativity via (fully 4-dimensional) Regge
         calculus.\hfil}
\medskip

    Much current activity in numerical relativity is centered around
making predictions which can be tested by the proposed Laser
Interferometry Gravitational Observatory (LIGO).$^{1,2}$
There is a need to solve Einstein's equations numerically for many
physical situations which could give rise to gravitational waves, so
that data {}from LIGO can be interpreted, and used, if appropriate, as
evidence for the existence of black holes.  More generally, numerical
solutions of Einstein's equations are invaluable for the understanding
of astrophysical data, and for guidance as to what experiments to
undertake.

     Methods of solving Einstein's equations numerically include finite
difference schemes and finite element schemes. Regge calculus is a type of
finite element method, and in this paper we shall describe a way of casting
it into the form of a highly efficient tool of numerical relativity.

     The basic idea of Regge calculus is the division of spacetime into
simplicial cells with flat interior geometry.$^3$ The dynamical variables
are the edge-lengths of the simplices, and the curvature, which is
restricted to the ``hinge simplices'' of codimension two, can be expressed
in terms of the defect angles at these hinges, where the flat cells meet.
Variation of the action leads to the simplicial form of Einstein's
equations.  The convergence of the Regge action and equations to the
corresponding continuum quantities has been investigated thoroughly and has
been shown to be satisfactory under quite general conditions.$^{4-9,22,30}$
Regge calculus has been applied to a large variety of problems in classical
and quantum gravity (see Ref.~10 for a review).  Numerical applications in
3+1 dimensions have been mainly to problems with symmetry and no general
code has been developed.$^{11-15,37}$ This is also the case in the
alternative approach known as null-strut calculus, which builds a
spacelike-foliated spacetime with the maximal number of null
edges.$^{16-20}$ Although null-strut calculus was used first to demonstrate
numerically the approximate diffeomorphism freedom in Regge
calculus,$^{18}$ and, except for the work described in ref.~30, was the
first fully 3+1 dimensional numerical scheme implemented $^{21}$, we have
not found a way to adapt it to the evolution scheme described in this
paper.  Therefore, unless we find an alternative decoupling scheme the
standard approach to Regge calculus appears to be much more tractable
numerically.

   In this paper, we describe how the implementation of ideas developed
almost twenty years ago$^{22}$ results in a much more efficient way of
using Regge calculus in numerical relativity.  The evolution of a spacelike
hypersurface can be achieved one vertex at a time, or in parallel for
vertices which are not connected by an edge.  This scheme is described in
Sec.~2, and the role of the Bianchi identities and the resulting freedom to
specify lapse and shift information are discussed in Sec.~3.  In Sec.~4, it
is shown how the scheme works for one of the lattices originally suggested
by Sorkin, and in Sec.~5 it is shown how it is related to previous
evolution schemes using Regge calculus.  Sec.~6 consists of a numerical
example, and Sec.~7 contains some concluding remarks.

\bigskip
\noindent
\line{\bf 2. General description of the evolution scheme.\hfil}
\medskip

     In a sense, the scheme which we will describe now is not a new
way of doing Regge calculus, but rather a new understanding of the
elegant way in which the standard evolution scheme works in Regge
calculus.  It is based on what is described by Sorkin for two
particular lattices;$^{22}$ the general validity of these ideas was
realized only recently and is summarized by Tuckey.$^{23}$

     Consider a time evolution problem in Regge calculus, and suppose that
all the edge-lengths of a solution, up to and including a given
triangulated spacelike hypersurface are known. Then, with an appropriate
continuation of the triangulation into the future, any vertex on that
surface can be evolved forward to the corresponding vertex on the next
hypersurface by solving only a small set of equations {}from the immediate
neighborhood of the vertex.

     To see how this works, look first at what happens in 2+1
dimensions.  Consider part of the two dimensional spacelike surface
surrounding one vertex (the star of that vertex) (Fig.~1).  To advance
this vertex to the next hypersurface, we introduce a new vertex
``above'' it; we connect this to the chosen vertex by a ``vertical
edge,'' and by ``diagonal'' edges to all the vertices on the original
surface to which the chosen vertex was connected.  The result of this
is to stick a tetrahedron, with apex at the new vertex, on each of the
triangles surrounding the original vertex (Fig.~2). It is rather like
erecting a tent above the chosen vertex, with the vertical edge as the
tent pole and the star (within the spacelike hypersurface) of the
chosen vertex as the floor of the tent.

     In order to evolve the new vertex, we need to find the lengths of
the new edges.  Recall that the empty space Regge equation for an edge
$L_i$,
$$
\sum_h {\partial A_h\over \partial L_i} \epsilon_h = 0 \ ,
\eqno{(2.1)}
$$
where the sum is over hinges $h$, with $A_h$ the volume content of the
simplicial hinge, and $\epsilon_h$ the defect angle there, involves only
the edge lengths of the simplices containing $L_i$.  Thus the only
equations which involve the new edges and edges already known are those for
the new vertical edge and for the edges in the original surface which
radiate {}from the chosen vertex.  Since by construction there is one such
edge corresponding to each new diagonal edge, there are precisely the same
number of equations as unknown edge lengths.  Thus one might think that we
could solve directly for the edges.  However, as we shall explain, the
equations are expected to have an approximate functional dependence among
them, which means that (for sufficiently fine triangulations) it will
probably be better to use the effective lack of determination to choose
lapse and shift, as in the continuum.

The construction just described can be generalized immediately to 3+1
dimensions.  We choose a vertex on the given spacelike hypersurface,
introduce a new vertex above it and connect the new vertex by a
``vertical'' edge to the chosen vertex and by ``diagonal'' edges to all the
vertices in the original hypersurface to which the chosen vertex was
joined.  Each tetrahedron in the original surface which contains the chosen
vertex now has based on it a 4-simplex, with apex at the new vertex.  Note
that again in this tent--like construction (Fig.~3), there is one diagonal
corresponding to each edge in the original surface radiating {}from the
chosen vertex.  We now use the variational equations for these edges in the
original surface and for the vertical edge; the only unknown edges which
these equations involve are the new vertical edge and the diagonal edges
and there is the same number of equations as unknowns, so in principle we
could again solve uniquely for the unknown edges. However in practice we
shall again ignore some of the equations and instead specify conditions on
the lapse and shift.  (Although fundamentally all the variational equations
have the same status, they play different roles in the chosen advancement
scheme.  Those associated with the ``vertical'' edges involve only the
region between successive spacelike hypersurfaces and are in that sense
constraint--type equations, whereas those associated with the other edges
stretch between three such hypersurfaces and are in that sense
evolution--type equations.  Nevertheless, eliminating 3 degrees of freedom
in favor of freely specifiable shift information effectively treats 3
combinations of the latter equations as ``constraint-like'' as well.)

So far, we have described how to advance just one vertex in time.  The
method can clearly be used to evolve the entire hypersurface, by advancing
the vertices one-by-one (see Fig.~4 for a representation of this process in
1+1 dimensions).  This process is completely general; it can be used for
any triangulation of a hypersurface
having any topology.

    Advancing the vertices one-by-one will not ordinarily be the most
efficient way of evolving a hypersurface.  If any two vertices in a
hypersurface are not connected by an edge, then they can be evolved to
the next surface at the same time without interfering with each other,
as shown schematically in Fig.~5.  Thus the method is obviously
parallelizable.

    Whichever the order in which the vertices are advanced, each
vertex has a unique predecessor and successor, and so the structure
always contains lines through it connecting the vertices in this way.
Thus it is ``washing line topology.''  The particular order of
advancement chosen hangs the washing on the line in a particular way.

   After a vertex has been evolved one step, the new exposed
3-surface, which is made up of most of the old hypersurface plus the
exposed tetrahedral faces of the 4-simplices which were added, has
the same triangulation as the original hypersurface.  This is because
the way that the 4-simplices are constructed ensures that each
tetrahedron in the original surface has a corresponding tetrahedron in
the new surface.  Thus the structure of the exposed 3-surface is the
same at all stages of the time evolution calculation, even though the
structure of the 4-dimensional slice  beneath that 3-surface depends on the
order in which the vertices were advanced to it.  This beautifully
simple structure could however be seen as a limitation if one wants to
allow less ``orderly" geometries where the triangulation or even the
topology changes in time. One might wish to generalize the scheme to
something more flexible where two washing lines knot together or one
branches out into two. It is possible that this could be achieved by using
a sequence of elementary moves on the triangulation.$^{24}$

     In a time evolution calculation the ``vertical'' edges always go
between hypersurfaces.  However the ``diagonal'' edges initially lie in a
new hypersurface but later may be regarded as going between surfaces when
more of the original surface is evolved (See Fig.~5).  What restriction
does causality place on them?  Consider the ``tent'' above a chosen vertex;
we are trying to calculate the 4-geometry within it {}from data given on
(and below) the base of the tent, so the tent as a whole ought to lie
within the domain of dependence of its base.  Even though the Regge
equations are in one sense being solved ``implicitly'' we expect that for
the solution to be stable, this causality condition should be satisfied,
i.e. the whole of the tent should be contained within the ``light pyramid''
on its spacelike base (Fig.~6).  In particular, the diagonal edges, which
form the boundary of the tent, must be spacelike.  On the other hand, the
vertical edge could in principle be timelike, null or spacelike.  Since it
seems that the diagonal edges will have to be spacelike, we see that all
the 3-surfaces, even the ones which could be thought of as intermediate
stages in the evolution of some starting hypersurface to the next, will
actually be spacelike.  A picture emerges of a whole pile of
``concertina'd'' spacelike 3-surfaces lying on top of each other in a lot
of places, with occasional gaps. To use yet another metaphor, it looks like
puff pastry.

The requirement that the diagonal edges be spacelike indicates that it
will not be a good idea to advance one particular vertex by several
steps while not evolving the surrounding ones. In that case, the
diagonal edges would be likely to become timelike rather soon, and
long narrow triangles would be produced, which have proven not to be
good in numerical calculations,$^{17}$ and are also known to be bad
for convergence.$^{38}$

Finally let us mention a possible application of these same ideas to the
initial value problem.  Since in the formulation just described there is no
meaningful distinction between initial value and time evolution equations,
one could contemplate building up an initial set of edge-lengths  by
the same process of ``advancing vertices'' described above in the context of
time evolution.  The only difference would be that certain edge-lengths
which occurred in the equations would now be unknown, and therefore freely
specifiable as initial data, rather than given by the results of prior
evolution.  This would seem to lead to a parallelizable method for solving
the initial value equations as well.  On the other hand, one would not
expect this to be possible, since the initial value problem has a
fundamentally elliptic character in contrast to the time evolution problem,
which in the continuum is hyperbolic, and therefore effectively local.
Perhaps the resolution is that an initial value solution produced by any
such scheme of ``advancing'' individual vertices would behave unstably, but
at any rate it seems a question worthy of further study.

\bigskip
\noindent
\line{\bf 3. Bianchi Identities and Lapse and Shift.\hfil}
\medskip

   Let us look now at the counting involved in the type of evolution
scheme just described.  At first sight there appears to be an exact
match at each vertex between the number of unknown variables
(edge-lengths) and the number of equations available to determine
these.  Two obvious and related questions arise. Firstly, are all the
equations independent? Secondly, what has become of the freedom to
choose lapse functions and shift vectors which exists in the continuum
theory?

   This matching of the numbers of unknowns and equations precisely
mirrors the continuum theory where we have ten unknown metric
components ($g_{\mu\nu}$) and ten Einstein equations ($G_{\mu\nu} = 8
\pi T_{\mu\nu}$) available to determine them.  However, we know that
in the continuum not all of the equations are independent.  There is a
four-fold redundancy corresponding to the contracted Bianchi identities
($\nabla \cdot G=0$).  Thus in the continuum theory we are free to impose
four conditions per point on the metric (one lapse condition and three
shift conditions) when evolving {}from one 3-geometry to the next.
Furthermore, the constraint equations once satisfied on the initial
3-geometry will automatically remain solved in the evolution.  This is
analogous to conservation of energy and momentum.  Therefore in practice we
must solve six evolution equations for the six metric parameters that
remain.

These continuum relationships have their counterparts in Regge
calculus and in particular when applied to the scheme described in the
previous section. In Regge calculus, the ordinary Bianchi identities
correspond (in four dimensions) to the statement that the product of
the rotation matrices for the hinges meeting on any edge is the
identity transformation.$^3$ In the limit of small defect angles, it
can be shown that this is equivalent to the usual continuum Bianchi
identities. In this case the equivalent of the contracted Bianchi
identity is that the sum over all edges meeting at a vertex, of the
equivalent of the Einstein tensor along that edge, is zero.$^{7,25}$
However, this vector equation is an approximate identity; it is only
exact in the linearized case or in the continuum limit. Thus it
provides four {\it approximate} equations per point, relating the
Regge equations for the edges at that vertex.  This is associated with
an approximate symmetry of the theory.

The implications of the contracted Bianchi identities for our evolution
scheme therefore go as follows.  When we evolve a vertex we will have $N$
equations for $N$ unknown edge-lengths.  Because of the approximate
four-fold redundancy per vertex we expect to be able to use $N-4$ of the
equations together with four externally imposed lapse and shift conditions
to solve for the unknown edges. For example, we might simply choose the
vertical edge-length as the lapse degree of freedom, and three of the
diagonal edges as the shift, and ignore the corresponding Regge equations
associated with the vertical edge and the corresponding edges in the tent
``floor''.

In a study of the Kasner cosmology, we have in fact imposed more
sophisticated lapse and shift conditions that involve four algebraic
relations among the $N$ variables corresponding to the standard
definitions of lapse and shift.$^{26,29}$ We have observed numerically
that the conservation of energy-momentum (in the sense of automatic
satisfaction of the unimposed equations) improves when we refine the
lattice, and that the Jacobian of second partial derivatives of the
action becomes more and more singular as we approach the continuum
limit, in agreement with earlier theoretical
predictions.$^{7,22,25,29,30,35,39,40}$

The approach developed here for evolution is in may ways similar to
the finite difference algorithm used by Kurki-Suonio, Laguna and
Matzner;$^{27}$ however, our discretized equations are written in an
implicit form.  To our knowledge this is the first implicit numerical
scheme in (3+1)-dimensional numerical relativity.

\bigskip
\noindent
\line{\bf 4. Sorkin's Lattice.\hfil}
\medskip

     We now illustrate the scheme by looking at one of the lattices for
which it was originally suggested.$^{22}$ Consider first the 2-dimensional
version, for a triangulation of $S^1\times R$.  We label vertices by $[t]$
or $[t^*]$ with $ t\in Z$, and join them according to the following rules:
$[t_1]$, $[t_2]$ or $[t_1^*]$, $[t_2^*]$ are joined if $0 <\mid t_1-t_2\mid
\leq 2$; and $[t_1]$, $[t_2^*]$ are joined if $0 < \mid t_1-t_2\mid \leq
1$.  The resulting lattice is shown in Fig.~7.  The triangles are of the
form $[t, t+1, t+2]$ where $t$ and $t+2$ were both starred or unstarred,
and $t+1$ may or may not be starred.  Clearly all vertices are equivalent.
What is happening here is that $S^1$ is being represented by a square, and
pairs of opposite vertices (e.g. $[0][0^*]$ and $[1][1^*]$) are staggered
in time.

   In four dimensions, we consider a triangulation of $S^3\times R$ in
which $S^3$ is tessellated by $\beta_4$, which has 8 vertices and 16
tetrahedra.  (More generally, one can use the family of posets depicted in
figure 5 of reference 36 to produce analogous triangulations for all the
$n$-spheres $S^n$, each such triangulation being the so-called
``order-complex'' of the corresponding poset.  The square is the $n=1$
reprsentative of this family and $\beta_4$ is the $n=3$ representative.)
The vertices are labeled as in the 2-dimensional case and the rules for
joining them are: $[t_1]$, $[t_2]$ or $[t_1^*]$, $[t_2^*]$ are joined if $0
<\mid t_1-t_2\mid \leq 4$; and $[t_1]$, $[t_2^*]$ are joined if $0 < \mid
t_1-t_2\mid \leq 3$. Again all vertices are equivalent, and we may regard
them as a succession of pairs $[0][0^*]$, $[1][1^*]$, $[2][2^*], \ldots$
. These pairs are the antipodal vertices of $\beta_4$ which in some sense
are staggered in time.  The 4-simplices are of the form
$[t,t+1,t+2,t+3,t+4]$ where either $t$ and $t+4$ are both starred or both
unstarred.  Each intermediate vertex may be starred or unstarred.

   The time evolution for this lattice proceeds as follows. We suppose that
all lengths up to and including vertices $[3]$ and $[3^*]$ are known. Now
consider vertex $[4]$, which is the vertex ``above'' $[0]$. $[4]$ is joined
to seven of the ``earlier'' vertices by the edges going ``backwards''
{}from it,
$$
[04][14][1^*4][24][2^*4][34][3^*4].
$$
These are the new edges in the tent above vertex $[0]$, the first of them
being the vertical edge and the others being the diagonal edges of this
tent.  Then consider the ``forward''--going edges {}from vertex $[0]$:
namely
$$
[01][01^*][02][02^*] [03][03^*][04],
$$
of which the first six are the horizontal edges of the tent and the last is
the vertical edge.  We can check that the Regge equations for these edges
involve only edges known already, plus the new edges listed above.  Thus we
have seven equations for seven unknowns, and if wanted to treat the
equations as truly independent, we could solve uniquely for the new
edge-lengths.  By doing so we would be letting the equations ``choose their
own gauge,'' and it is conceivable that they would make a reasonable
choice, since we are rather far {}from the continuum limit.  If, on the
contrary, the approximate symmetries in Regge calculus manifested
themselves in an unstable behavior of the resulting solutions, then we
could instead eliminate 4 of the seven equations in favor of freely
specifiable lapse and shift data, as discussed above. With either
procedure we would, in the language of Sec.~2, have chosen vertex $[0]$ and
advanced it to vertex $[4]$.

     Notice that the procedure for advancing to vertex $[4]$ involves
neither vertices $[0^*]$ or $[4^*]$ so in fact we could carry through
the advance of vertex $[0^*]$ to vertex $[4^*]$ at exactly the same
time.  This is because the vertices $[t]$ and $[t^*]$ have no edge in
common.

     We can now repeat the process to find everything up to and including
vertices $[5]$ and $[5^*]$ and so on.  At each step we will have two sets
of 7 equations in 7 unknowns to solve (each set containing three
``dynamical relations'' and four ``gauge conditions'').

     Before performing the time evolution, we must solve the initial value
problem. The equations used in advancing to vertices $[4]$ and $[4^*]$
involve edges going back as far as $[-3]$ and $[-3^*]$, so for initial
data, we need to know the lengths of all the edges between $[-3]$, $[-3^*]$
and $[3]$, $[3^*]$. This involves a total of 66 edges. Amongst these are
just 18 edges whose Regge equations involve only edges {}from the 66. Thus
one way of proceeding is to specify freely 48 edges and then solve the 18
equations for the remaining ones.

\bigskip
\noindent
\line{\bf 5. Comparison with other numerical work in 3+1 Regge calculus.
\hfil}
\bigskip

  Before attempting to relate other approaches to that of this paper,
let us emphasize the differences. Typical of earlier 3+1 approaches
are those of Collins and Williams,$^{11}$ Porter$^{13,14}$ and
Brewin.$^{33}$ The basic idea is to take successive spacelike
hypersurfaces triangulated by tetrahedra, and to join corresponding
vertices by timelike lines, thus constructing a set of 4--prisms which
tessellate the spacetime. The shape of a 4--prism is not uniquely
determined by its edge lengths, so it is necessary to give further
information to eliminate the floppiness. In the study of the Friedmann
universe by Collins and Williams,$^{11}$ there is sufficient symmetry
to determine the shapes of the prisms. In the work of Porter$^{13,14}$
and later of Dubal,$^{28}$ angles in the faces of the prisms are
introduced as extra variables related to the extrinsic curvature. In
all of these approaches, the Regge equations fall into two categories,
evolution equations which arise {}from variation of spacelike edges, and
constraint equations which come {}from variation of timelike edges. The
equations are coupled together in such a manner as to make local solution
impossible.

   Apart {}from the different triangulations and the appearance of distinct
categories of equations, these methods also differ {}from the Sorkin
approach in that the spacelike hypersurfaces of a distinguished family are
clearly displayed at different times, whereas those in the Sorkin method
described in Sec.~2 are represented as staggered in time, in the manner of
puff-pastry. In the latter method, a subset of the surfaces may be
identified with those of the former method (though in general the choice of
subset is not unique).

   We now describe how the prism construction can be modified so that the
general method of Sec.~2 is applicable to it and then describe the
relationship in detail for the particular lattice of Section 4 above.

   The modification required is a very simple one; all that is required is
the addition of enough diagonals or braces to divide each prism into four
4--simplices. (We note with apology the possibility for confusion between
these ``diagonals'' and the ``diagonal edges'' discussed up until now. The
``diagonals'' here are defined with respect to a particular family of
non-intersecting 3-dimensional hypersurfaces. The ``diagonal edges''
discussed previously are defined only with respect to a particular
evolution step.) There are various prescriptions for doing this. One
possibility is to order all the vertices in the upper hypersurface (which
forms the top layer of all the prisms), and then connect the first vertex
in each upper simplex (for all simplices of all dimensions) to every vertex
of the corresponding simplex in the lower hypersurface. Then the
$k$--simplices are just the subsets of a single cell spanned by vertices
all of which are connected together. The resulting simplicial complex is
then exactly what would be obtained by performing the evolution process of
Sec.~2 on the lower hypersurface, advancing the vertices in exactly the
order imposed on the upper hypersurface. (A similar construction is
possible in any dimension, see for example ref.~41 and section 3 of
ref.~36.)

   Note that what we are claiming here is not that the prism methods are
equivalent to that of Sec.~2, but rather that the lattices used can be
modified to fit the new method. The prism methods per se are quite distinct
dynamically, involving for example the use of angles as variables. Once the
diagonal edges have been introduced, there is no point in using any extra
variables, since one returns to the situation in conventional Regge
calculus where the edge lengths completely specify the geometry.  Clearly,
the curvature in the 4-prism based geometries is less general than that
allowed after the subdivision into 4-simplices, since in the prism approach
the curvature is constant across a 2-prism, which is not true when it is
split into 2 triangles.

   Let us now show how to introduce the diagonals into the prism
construction for a particular lattice, which reproduces the lattice
described in Sec.~4.  Consider $\beta_4$, the 16--cell tessellation of
$S^3$. Label the vertices at one moment of time $[0], [1], [2], [3], [0^*],
[1^*], [2^*], [3^*]$ (to tie in with Sorkin's notation), as shown in
Fig.~8. At the next moment of ``time'', label the vertices $[4], [5], [6],
[7], [4^*], [5^*], [6^*], [7^*]$. We now join corresponding vertices (those
differing by 4) at successive times to obtain a collection of 4--prisms
which we now divide into 4--simplices in the following way. In each
quadrilateral face within a prism, we draw in a diagonal joining the two
vertices with nearer numbers. For example in the prism with lower
tetrahedral face $[0^*1^*23]$ and upper tetrahedral face $[4^*5^*67]$
(Fig.~9), we insert the diagonals $[1^*4^*]$, $[24^*]$, $[25^*]$, $[34^*]$,
$[35^*]$ and $[36]$. The prism is now divided into 4--simplices
$[0^*1^*234^*]$, $[1^*234^*5^*]$, $[234^*5^*6]$ and $[34^*5^*67]$. It is
clear that the rule by which vertices are joined produces exactly the same
lattice as in the Sorkin case. This means that when we apply the method of
Sec.~2, the dynamics of the evolution procedure are exactly the same; to
evolve to the next hypersurface we have to solve two sets of seven
equations for seven unknowns, four times over. The interpretation of the
procedure may be slightly different here, as already pointed out, in that
there is no suggestion of the designated spacelike hypersurfaces being
staggered in time or concertina'd in shape.

    The initial value problem here differs slightly {}from the Sorkin
case. In order to evolve the spacetime, we need to specify two spacelike
slices and the spacetime between them. This involves 24 spacelike edges in
each hypersurface, plus 8 timelike edges and 24 diagonals in between,
making a total of 80 edges. The equations involving only these edges are
those for the timelike edges and diagonals, a total of 32 equations. Thus
we may specify freely 48 edges (in exact agreement with the Sorkin case)
and use the available equations to solve for the other 32. The difference
arises because the initial value problem identified here corresponds to the
minimal initial value problem identified in Sec.~2, together with the
addition of the two evolution steps of vertex [3] to [7] and [$3^*$] to
[$7^*$], which provides the additional 14 equations and 14 unknowns.

   One of the main points of this section was to show that local evolution
(and sometimes parallel evolution) is possible, and indeed natural, in the
previous work on 3+1 Regge calculus using prisms, provided the prisms are
divided into 4--simplices by the insertion of extra edges. The possibility
of advancing several vertices at the same time depends on the symmetry of
the lattice. We have already seen how this happens for the lattice based on
$\beta_4$, and we shall end this section by looking briefly at how it works
for the other regular tessellations of $S^3$.

   For $\alpha_4$, the 5--cell triangulation of $S^3$, there are 5 vertices
and 10 edges.  In this case, since all the vertices are connected to each
other, it is not possible to advance any pair simultaneously. The evolution
procedure can be shown to involve solving 5 equations for 5 unknowns, 5
times over, to advance to the next spacelike hypersurface.

The 600-cell triangulation of $S^3$ can be built up {}from 30 blocks of 20
tetrahedra meeting at single vertex. Clearly these blocks meet the
condition that none of the central vertices shares an edge, and so we can
advance the 30 central vertices at the same time. We then consider another
division into 30 blocks, and advance their central vertices
simultaneously. Since the total number of vertices is 120, the process will
have to be repeated 4 times to evolve a spacelike hypersurface to the next
distinct spacelike hypersurface. Now in each hypersurface, 12 edges meet at
each vertex; we use the Regge equations for those edges and for the
vertical edge {}from the vertex to its advancement in the next
hypersurface, to solve for that vertical edge and the 12 diagonal edges
involved. Thus the whole evolution process {}from one surface to the next
distinct surface involves solving four times 30 sets of 13 equations for 13
unknowns, making a total of 1560 edges to be solved for ultimately. These
are made up of 720 edges in the new spacelike hypersurface, 120 vertical
edges and 720 diagonals (which are also spacelike edges, playing the roles
first of diagonal edges and then horizontal edges in the evolution steps).

   It is interesting to note the relationship between the number of
variables to be solved for at each stage, and the number of cells in the
triangulation. The cases mentioned show the following progression:
$$
 \hbox{ tetrahedra:\ \ }5,16,600,\infty;
 \hbox{\hskip .5 truein}
 \hbox{unknowns:\ \ }5,7,13,15.
$$
Here the last entry is taken {}from Ref.~22 for the $R^4$ triangulation and
is the same number obtained in the asymptotic limit for the
``quantity--production'' lattice introduced in Ref.~17. (However, this by
no means proves that the average number of equations per vertex is
necessarily 15 in the limit of an arbitrarily large number of tetrahedra.
For example, the average number of edges per vertex for an infinitely
barycentrically--subdivided triangulated 3-dimensional hypersurface is
13/2, which leads in our evolution scheme to 13+1 = 14 equations per
vertex, rather than 15; and doing the subdivision directly on the
4-dimensional complex for comparison, would lead to the still smaller value
of 25/2 = 12.5 edges or Regge equations per vertex. On the other hand,
neither of these subdivision schemes is physically appropriate since
repeated barycentric subdivision produces infinitely squashed simplices in
the limit of infinite refinement.)

%
%

\bigskip
\noindent
\line{\bf 6. Preliminary Numerical Example.\hfil}
\medskip

We now illustrate the scheme described in this paper by applying it to the
evolution of the 600-cell tessellation of $S^3$ mentioned in the previous
section.  Matter will be included in the form of pressure--less dust, and
we will look for homogeneous solutions, so the model will be an
approximation to the Friedmann universe.$^{11,32}$

We label the edge lengths in a way appropriate to homogeneous solutions. We
distinguish a particular family of spacelike hypersurfaces, where each
surface is generated by evolving every vertex in the preceding surface
once. The (spacelike) edges lying in these surfaces are called spatial and
their lengths are denoted by $l_i$. The proper lengths of the (timelike)
vertical edges going between these hypersurfaces are denoted by $v_i$, and
the lengths of the (spacelike) diagonals between these surfaces are denoted
by $d_i$. In each case, $i$ labels the edges in the class.

Rather than considering the usual description of matter in the Friedmann
universe as dust of uniform density, we model it by 120 dust particles each
of mass $M/120$, where $M$ is the total mass, with one particle moving
along each of the vertical edges along which the vertices evolve. We note
that this entails a kinematical restriction on the particle paths with
respect to the geometry, but as the paths made up {}from the consecutive
vertical edges will be geodesics in our homogeneous solutions, the
particles will indeed move on geodesics. The Regge action is then (in units
such that $c = G = 1$)
$$
I = {1\over 8\pi}\sum_h A_h \epsilon_h - \sum_i {M\over 120} v_i \ ,
 \eqno{(6.1)}
$$
where the second sum is over vertical edges of proper length $v_i$.  The
Regge equations are obtained by varying this with respect to the
edge-lengths (cf.~Eq.~2.1).  Variation with respect to a vertical edge
length $v_i$ gives
$$
\sum_h {\partial A_h\over \partial v_i} \epsilon_h = {\pi M\over 15} \ ,
 \eqno{(6.2)}
$$
and with respect to a diagonal length $d_i$ or spatial edge length $l_i$
gives
$$
\sum_h {\partial A_h\over \partial d_i} \epsilon_h = 0 \ ,\ \ \
\sum_h {\partial A_h\over \partial l_i} \epsilon_h = 0 \ .
\eqno{(6.3)}
$$

We now outline the evolution procedure used here. As discussed in the
previous section, the general evolution step involves 13 equations in 13
unknowns, and is expected to allow the approximate freedom to impose four
lapse and shift conditions. We will however not consider this general
situation, but will instead assume that there exist homogeneous solutions,
i.e.~that there exist solutions where the lengths $l_i$ of the edges in any
hypersurface are equal, all $v_i$ going between any two adjacent
hypersurfaces are equal, and all $d_i$ going between any two adjacent
hypersurfaces are equal. We will then solve a minimal subset of equations
6.2 and 6.3 needed to generate a solution given this assumption. (In fact
we will only use equations of the type of 6.2.) The condition on the $v_i$
may be viewed as being merely a choice of lapse, but the conditions on the
$d_i$ and $l_i$ are certainly more than a choice of shift, and involve an
assumption about the dynamics of the system. Our approach may be compared
with that of Brewin$^{33}$, who made a similar assumption but substituted
it into the action in order to derive his equations, thus effectively
finding stationary points only with respect to this class of
geometries. His and our approaches are correct if and only if the full
equations admit such solutions.

To implement this approach we take some hypersurface, consider a vertex
$[0]$ in this hypersurface which is connected by spatial edges to the
twelve other vertices $[1],[2],\ldots,[12]$, and evolve $[0]$ to the new
vertex $[0^*]$. This gives 13 new edges, the vertical one $[00^*]$ and 12
diagonals. The Regge equations available are one for the vertical edge
$[00^*]$ and 12 for the spatial edges $[01],[02],\ldots,[0\,12]$.  (For the
first evolution step, the latter 12 equations are not really available
since they involve edges in the previous spacetime slice which have not
been specified. However, when the process is repeated after the first
evolution of all vertices, these equations could be used.) We then assume
all spatial edge lengths $l_i$ in the surface are equal to some (known)
$l_0$, we set the proper length of the vertical edge $[00^*]$ equal to some
$v_0$ (i.e.~we choose the lapse), and we assume that the twelve diagonals
all have the same length, denoted by $d_0$. We then use the single Regge
equation for the vertical edge $[00^*]$ to solve for $d_0$.  There are 12
identical triangular hinges, with edge-lengths $l_0$, $v_0$ and $d_0$ and
area $A_{010^*}$, sharing $[00^*]$, and 5 identical 4-simplices meeting on
each triangle, so the Regge equation is
$$
v_0 \cot{\theta_0} \epsilon_{010^*} = {\pi M\over 90} \ ,
 \eqno{(6.4)}
$$
where
$$
\cot{\theta_0} = {d_0^2+l_0^2-v_0^2\over 4 A_{010^*}} \ ,
\eqno{(6.5)}
$$
and
$$
\epsilon_{010^*} = 2\pi - 5 \theta_{010^*} \ ,
\eqno{(6.6)}
$$
with $\theta_{010^*}$ being the hyperdihedral angle at $[010^*]$ in
$[01230^*]$ say.

Now consider the evolution of vertex $[1]$ to $[1^*]$. This gives the new
vertical edge $[11^*]$, a new spatial edge $[0^*1^*]$, and 11 new
diagonals. The Regge equations are one vertical $[11^*]$, one diagonal
$[10^*]$, and 11 spatial (the latter 11 not being available the first time
around). We again use the lapse freedom to set the length of the vertical
edge to be $v_0$ (consistent with homogeneity), and we assume that all new
diagonals have the length $d_0$ (now known). We then solve the Regge
equation for the vertical edge $[11^*]$ to find the length of the new
spatial edge $[0^*1^*]$, denoted by $l_1$. In this case there are three
types of triangles meeting on $[11^*]$ and two types of 4-simplex, and the
Regge equation reduces to
$$
 v_0\left[11 \cot{\theta_0} \epsilon_{121^*} +
 \cot{\theta_1} \epsilon_{10^*1^*}\right] = {2\pi M\over 15} \ ,
 \eqno{(6.7)}
$$
where
$$
\cot{\theta_1} = {d_0^2+l_1^2-v_0^2\over 4 A_{10^*1^*}} \ ,
\eqno{(6.8)}
$$
$$
\epsilon_{121^*} = 2\pi - 2\theta_{010^*} - 3\theta_{121^*} \ ,
\eqno{(6.9)}
$$
and
$$
\epsilon_{10^*1^*} = 2\pi - 5\theta_{10^*1^*} \ ,
\eqno{(6.10)}
$$
with $A_{10^*1^*}$ being the area of triangle $[10^*1^*]$,
$\theta_{121^*}$ the hyperdihedral angle at $[121^*]$ in $[1230^*1^*]$
and $\theta_{10^*1^*}$ the hyperdihedral angle at $[10^*1^*]$ in
4-simplex $[1230^*1^*]$.

Under our assumption of homogeneity, this completes the evolution {}from the
original hypersurface to the new one, since all vertical edges between the
surfaces must have length $v_0$, all diagonals have length $d_0$, and all
spatial edges on the new surface must have length $l_1$. The evolution to
the next surface may then be carried out in the same fashion, i.e.~the next
diagonal length $d_1$ is found {}from the vertical equation associated with
the evolution of one vertex (given a choice $v_1$ of the next vertical edge
length), and the next spatial length $l_2$ is found {}from the evolution of
an adjacent vertex. Thus the data required to specify a solution of the
form we assume are the length $l_0$ of the spatial edges on some initial
surface, and the vertical length between each pair of consecutive surfaces.
In our calculations, the vertical lengths between all consecutive surfaces
were chosen to be the same, equal to $v$.

We will discuss the space of solutions in general below, but we commence
with those which correspond to the continuum (Friedmann) solution,
illustrated by the one shown in figure 10. For the purposes of this figure,
the spatial edge length $l$ on each hypersurface is converted into an
equivalent 3-sphere radius $a$ by equating the volume of the simplicial
hypersurface with that of a smooth 3-sphere, i.e.
$$
2\pi^2 a^3 = 600{l^3\over 6\sqrt 2} \ ,
\eqno(6.11)
$$
and this is compared with the scale factor of the Friedmann universe as a
function of proper time.  (The other variable one might want to plot would
be the diagonal length $d$, but in the present case $d$ is virtually
identical in magnitude to $l$, due to the small value chosen for the lapse
parameter $v$.)  Note that the proper time elapsed between two consecutive
simplicial hypersurfaces may be defined in different ways. In figure 10 we
have used the proper time $\delta t$ as measured along a geodesic running
{}from the centre of a tetrahedron in one surface to the centre of the
corresponding tetrahedron in the neighbouring surface, which is given by
(see ref.~11, equation 43)
$$
\delta t = \left(v^2 + {3\over 8}(l'-l)^2\right)^{\!1/2} \ ,
\eqno(6.12)
$$
where $l$ and $l'$ are the spatial edge lengths on the two surfaces. This
gives a slightly slower evolution of the approximate solution as a function
of proper time than does the proper time elapsed along a vertical edge,
which is just equal to $v$.

The solution shown in figure 10 is a time-symmetric solution of the Regge
equations, in which the maximum spatial edge length occurs on two
consecutive surfaces, and the other surfaces are symmetrical about these
two. That is, if the central surfaces are numbered $0$ and $1$, then pairs
of surfaces numbered $-n$ and $1+n$, $n$ a positive integer, have the same
spatial edge lengths. (The diagonal lengths $d_i$ are also symmetrical
about the central value of $d_i$ going between surfaces 0 and 1.)
For this solution it is found that the maximum spatial edge length
$l$ satisfies
$$
12\,l\left(2\pi - 5\cos^{-1}\left({1\over 3}\right)\right) =
16\pi{M\over 120} \ .
\eqno(6.13)$$
This is the Regge analogue of the equation $^3\!R = 16\pi\rho$ (see
ref.~32), which of course is the constraint equation applying at the
surface of time symmetry in the continuum solution. {}From 6.11 and 6.13
we find that for the same mass $M$, the equivalent maximum $a$ of the Regge
solution is slightly less than the maximum scale factor $a = 4M/(3\pi)$ of
the Friedmann solution.  In figure 10 we have compared our Regge solution
($M = 10.2$) with the Friedmann solution ($M=10.0$) of equal maximum radius
$a$.

The Regge equations have a second time-symmetric solution, in which the
maximum spatial length $l$ is reached on a single hypersurface, and the
other hypersurfaces are symmetric in pairs around this hypersurface. This
maximum $l$ is slightly larger than that given by equation 6.13. There are
also solutions which have a maximum $l$ whose value lies between these two,
which is reached on one hypersurface, and which are not time-symmetric.
These are all acceptable models of the Friedmann universe, and on the scale
of figure 10 they are indistinguishable {}from the solution shown there.

The solution in figure 10 has the vertical edge length $v = .0102$. The
separate plotted points overlap for most of the solution but may be
distinguished toward the ends of the evolution. For $v=.102$, the points in
the solution are of course spaced further apart, but they lie on the same
curve as a function of proper time. This supports the expectation that $v$
is a gauge degree of freedom.

As was observed previously by Brewin$^{33}$ for his models, the evolution
of the Regge universe stops at a finite volume, i.e.~solutions cease to
exist before the universe collapses to zero volume.  Presumably this
corresponds to the collapse becoming so fast that the vertical edges would
have to become spacelike in order for solutions to exist, but we have not
investigated this further. The endpoint is not very sensitive to the
value of $v$ used in the calculations.

The Regge equations also have solutions which do not correspond to the
Friedmann universe. To begin with, equation 6.7 has two roots for $l_1$
for given values of $l_0$, $d_0$ and $v_0$. One of these roots
satisfies the equation
$$
d_0^2 = -v_0^2 + l_0 l_1 \ ,
\eqno(6.14)$$
which is the same as equation 6.6 of ref.~33. Brewin obtained this equation
{}from the assumption that the 4-simplices in the model are grouped into
regular, untwisted 4-prisms. We have not investigated the geometrical
interpretation of the other root. The root of equation 6.7 satisfying 6.14
was the one chosen in obtaining the solution shown in figure 10. The other
root for $l_1$ is larger in the expansion phase and smaller in the
contraction phase, giving a solution which evolves about twice as fast as
this one.

For given values of $l_0$ and $v_0$, equation 6.4 typically has four roots
for $d_0$, which may be naturally grouped into pairs, with one member
{}from each pair giving an expansion of the universe, and the other giving
a contraction. One of these pairs of roots leads to the expanding and
recontracting solution shown in figure 10, while the other gives a universe
which expands without limit {}from a finite initial volume, plus the
corresponding time-reversed solution. The physically interesting roots lie
in a sharp valley in the centre of a peak of the function on the left hand
side of equation 6.4 (considered as a function of $d_0$), and could not be
reliably found with a Newton-Raphson algorithm, so the bisection method was
used. Other spurious solutions are allowed in which we jump {}from one pair
of roots to another, or in which the expansion/contraction of the universe
reverses arbitrarily. Some of these spurious solutions clearly arise
because equation 6.4 is associated with a vertical edge, and so is not
sensitive to data prior to the current hypersurface.  In particular, an
expansion-contraction ambiguity had to be present for this reason.  Thus
consideration of the equations ignored in our analysis can be expected to
eliminate some or all of the spurious solutions.

In summary, we have obtained a solution which is a reasonable approximation
to the Friedmann universe, and which compares acceptably to previous
solutions$^{11,33}$. We emphasise that our method generalises in a
straightforward way to more complicated models, with only a moderate
increase in the size of the individual algebraic problems to be solved.
The next stage in testing this method is to impose the homogeneity
condition on the initial data, but only to impose lapse and shift choices
on the evolution, and to verify that the full equations do admit the
homogeneous solutions discussed here, while excluding the spurious
solutions which our truncated treatment permitted.  As a first step, one
can check how well the solution reported here satisfies the equations which
were ignored in obtaining it.  Work on this is in progress, together with
further investigations of the 5--, 16-- and 600--cell universes, as well as
work on the evolution of the Kasner universe using a general code for $T^3$
topologies based on the ``quantity--production lattice.''$^{34,17}$ Results
will be reported elsewhere.

\bigskip
\noindent
\line{\bf 7. Conclusions.\hfil}
\medskip

We have described an implicit evolution scheme for Regge calculus. The most
important feature of this scheme is the possibility of local evolution for
one vertex at a time, a result of the fact that the Regge equation for an
edge involves only those simplices containing the edge. Parallel evolution
is possible for groups of vertices with no edges in common. Both of these
features are very important for numerical calculations.

An important question is whether all the variational equations are
independent.  In the continuum limit, as we have discussed, one expects the
approximate symmetries in the lattice to become exact, giving rise to exact
contracted Bianchi identities. The equations will no longer be independent,
and the freedom to specify the lapse and shift will be recovered. Away
{}from the continuum limit, there will be approximate contracted Bianchi
identities, giving four conditions per vertex.  It will thus probably be
best to ignore four equations per vertex and utilize the resulting freedom
to specify the discrete lapse and shift, as we did in Sec.~6 with the
lapse.  As for our results there, it has been shown in other particular
cases$^{29}$ as well, that the evolution is independent of the values of
the discrete lapse and shift to high accuracy, an echo of the
diffeomorphism symmetry of the continuum.

When we compare the new method with the modification of the prism--based
approach described earlier, we see that the symmetries of the lattice are
more obvious in the new approach than for the modified prism method, since
there is some arbitrariness in dividing the prisms into 4-simplices, and
since not all of the symmetries need take the prism-based hypersurfaces
into themselves.

The distinction between evolution equations (arising {}from variation of
spacelike edges) and constraint equations ({}from variation of timelike
edges and diagonals) in the prism approach is not present a priori in the
new method, although some distinction may arise as a result of
consideration of approximate symmetries, or {}from a choice of a
distinguished family of spacelike hypersurfaces.  In the prism approach, it
is clear which hypersurfaces are meant to correspond to constant times; in
the new method, the question of how the hypersurfaces are staggered in time
seems to depend on the initial data, which provides the only distinction in
the theory between the different classes of hypersurface.

The numerical example described in this paper illustrates the effectiveness
of this approach to Regge calculus.  Clearly there are many aspects still
to be explored; these include the introduction of more general source terms
(such as an electromagnetic field, to which our scheme extends essentially
unchanged$^{31}$), the construction of lattices with different topologies,
and the development of efficient ways to present the results. It seems at
this stage of the development of the theory that it has the attractive
features of finite difference methods and also appears to have the
advantage of being able to handle the topological complications found in
such problems as the two black hole collision and the spatially closed
cosmologies.  There is a need to test it on evolution problems with many
degrees of freedom and to compare its efficiency and accuracy with those of
the corresponding finite difference schemes.

\bigskip
\noindent
\line{\bf Acknowledgements.\hfil}
\medskip
The authors are grateful to I.~Pinto and J.~C.~Miller who organized the
very stimulating {\it Workshop on Numerical Applications of Regge Calculus
and Related Topics} in Amalfi in 1990 where this work was begun. We thank
Ian Drummond, Arkady Kheyfets, Pablo Laguna, John Porter and Martin Ro\v
cek for helpful discussions. WAM acknowledges support {}from the Air Force
Office of Scientific Research. RDS acknowledges support {}from the National
Science Foundation under Grant PHY--9307570. PAT is supported by a
fellowship {}from the Alexander von Humboldt Foundation. RMW thanks Los
Alamos National Laboratory for hospitality during the completion of this
work.

{\frenchspacing  

\bigskip
\noindent
\line{\bf References.\hfil}
\medskip
\item{[1]} K. S. Thorne, ``Gravitational radiation,'' in {\it
300 Years of Gravitation}, eds. S. Hawking and W. Israel (Cambridge
University Press, 1987) 330--458.
\smallskip
\item{[2]} K. S. Thorne, ``Sources of Gravitational Waves and
Prospects for their Detection,'' Caltech preprint GRP--234 (1990).
\smallskip
\item{[3]} T. Regge ``General relativity without
coordinates,'' {\it Nuovo Cimento} {\bf 19}, 558--571 (1961).
\smallskip
\item{[4]} J. Cheeger, W. M\" uller and R. Schrader, ``On the curvature
of piecewise flat spaces,'' {\it Comm. Math. Phys.} {\bf 92}, 405-54 (1984).
\smallskip
\item{[5]} R. Friedberg and T. D. Lee, ``Derivation of Regge's Action
{}from Einstein's Theory of General Relativity,'' {\it Nucl. Phys.}
{\bf B242}, 145-66 (1984).
\smallskip
\item{[6]} G. Feinberg, R. Friedberg, T. D. Lee and H. C. Ren, ``Lattice
Gravity Near the Continuum Limit,'' {\it Nucl. Phys.} {\bf B245}, 343-68
(1984).
\smallskip
\item{[7]} J. W. Barrett, ``The Einstein Tensor in Regge's
Discrete Gravity Theory,'' {\it Class. Quantum Grav.} {\bf 3}
203--6 (1986).
\smallskip
\item{[8]} J. W. Barrett, ``A convergence Result for Linearised Regge
Calculus,'' {\it Class. Quantum Grav} {\bf 5}, 1187-92 (1988); and
J. W. Barrett and R. M. Williams, ``The convergence of Lattice
Solutions of Linearised Regge Calculus,'' {\it Class. Quantum Grav.}
{\bf 5}, 1543-56 (1988).
\smallskip
\item{[9]} L. Brewin, ``Equivalence of the Regge and Einstein Equations
for Almost Flat Simplicial Spacetimes,'' {\it Gen. Rel. Grav.} {\bf 21},
565-83 (1989).
\smallskip
\item{[10]} R. M. Williams and P. A. Tuckey, ``Regge calculus: a
brief review and bibliography,'' {\it Class. Quantum Grav.} {\bf 9},
1409--22 (1992).
\smallskip
\item{[11]} P. A. Collins and R. M. Williams, ``Dynamics of the
Friedmann Universe Using Regge Calculus,'' {\it Phys. Rev.}  {\bf D7},
965--71 (1973).
\smallskip
\item{[12]} L. Brewin, {\it The Regge Calculus in Numerical Relativity},
Ph. D. Thesis Monash Univ. (1983).
\smallskip
\item{[13]} J. D. Porter, ``A New Approach to the Regge Calculus:
I. Formalism,'' {\it Class. Quantum Grav.} {\bf 4}, 375-89 (1987).
\smallskip
\item{[14]} J. D. Porter, ``A New Approach to the Regge Calculus:
II. Application to Spherically-Symmetric Vacuum Space-Times,''
{\it Class. Quantum Grav.} {\bf 4}, 391-410 (1987).
\smallskip
\item{[15]} M. R. Dubal, {\it Numerical Computations in General
Relativity}, Ph. D. Thesis SISSA Trieste (1987).
\smallskip
\item{[16]} W. A. Miller and J. A. Wheeler, ``4-Geodesy,''
{\it Nuovo Cimento} {\bf 8}, 418-34 (1985).
\smallskip
\item{[17]} W. A. Miller, ``Geometric Computation:
Null--Strut Geometrodynamics and the Inchworm Algorithm,'' in {\it
Dynamical Spacetimes and Numerical Relativity}, ed. J. Centrella
(Cambridge University Press, 1986) 256--303.
\smallskip
\item{[18]} A. Kheyfets, W. A. Miller and J. A. Wheeler,
``Null-Strut Calculus: The First Test,'' {\it Phys. Rev. Lett.},
{\bf61}, 2042-45 (1988).
\smallskip
\item{[19]} A. Kheyfets, N. J. LaFave and W. A. Miller,
``Null-Strut Calculus I: Kinematics,'' {\it Phys. Rev.} {\bf D41},
3628-36 (1990).
\smallskip
\item{[20]} A. Kheyfets, N. J. LaFave and W. A. Miller,
``Null-Strut Calculus II: Dynamics,'' {\it Phys. Rev.} {\bf D41},
3637-51 (1990).
\smallskip
\item{[21]} W. A. Miller developed a (3+1)-dimensional initial-value
code for null-strut calculus (Oppie-3) and applied it to the Kasner
cosmology, Unpublished (1990).
\smallskip
\item{[22]} R. D. Sorkin, ``Time--Evolution Problem in
Regge Calculus,'' {\it Phys. Rev.} {\bf D12}, 385-96 (1975).
\smallskip
\item{[23]} P. A. Tuckey, ``The Construction of Sorkin Triangulations,''
{\it Class. Quantum Grav.} {\bf 10}, L109--13 (1993).
\smallskip
\item{[24]} J. W. Barrett, ``A Mathematical Approach to Numerical
Relativity,'' in {\it Approaches to Numerical Relativity, Proc. of the
International Workshop on Numerical Relativity Southampton December 1991},
ed. R. d'Inverno (Cambridge: Cambridge University Press, 1992) 103--13.
\smallskip
\item{[25]} W. A. Miller, ``The Geometrodynamic Content of
the Regge Equations as Illuminated by the Boundary of a Boundary
Principle,'' {\it Found. Phys.} {\bf 16}, 143--69 (1986).
\smallskip
\item{[26]} C. Misner, K. S. Thorne and J. A. Wheeler, {\it Gravitation}
(W. H. Freeman and Co., 1973) 505--8.
\smallskip
\item{[27]} H. Kurki-Suonio, P. Laguna and R. A. Matzner, ``Inhomogeneous
Inflation: Numerical Evolution,'' {\it Phys. Rev.} {\bf D48}, 3611--24
(1993).
\smallskip
\item{[28]} M. R. Dubal, ``Relativistic Collapse Using Regge Calculus:
I. Spherical Collapse Equations,'' {\it Class. Quantum Grav.} {\bf 6},
1925-41 (1989).
\smallskip
\item{[29]} M. Galassi, ``Lapse and Shift in Regge Calculus,'' {\it Phys.
Rev.}  {\bf D47}, 3254--64 (1993).
\smallskip
\item{[30]} R. D. Sorkin,  {\it Development of Simplicial Methods for the
Metrical and Electromagnetic Fields}, Ph.D. Thesis California Institute of
Technology (1974) (available {}from University Microfilms, Ann Arbor,
Michigan).
\smallskip
\item{[31]} R. D. Sorkin, ``The Electromagnetic Field on a Simplicial Net,''
{\it J. Math. Phys.} {\bf 16}, 2432--40 (1975); and Erratum, {\it
J. Math. Phys.} {\bf 19}, 1800 (1978).
\smallskip
\item{[32]} J. A. Wheeler, ``Geometrodynamics and the Issue of the Final
State,'' in {\it Relativity, Groups and Topology}, eds. B. DeWitt and
C. DeWitt (New York: Gordon and Breach, 1964) 463--500.
\smallskip
\item{[33]} L. Brewin, ``Friedmann Cosmologies via the Regge Calculus,''
{\it Class. Quantum Grav.} {\bf 4}, 899--928 (1987).
\smallskip
\item{[34]} M. Galassi, Private Communication (1994).
\smallskip
\item{[35]} J. B. Hartle, ``Simplicial Minisuperspace I. General
Discussion,'' {\it J. Math. Phys.} {\bf 26}, 804--18 (1985); ``Simplicial
Minisuperspace II.  Some Classical Solutions on Simple Triangulations,''
{\it J. Math. Phys.} {\bf 27}, 287--95 (1986).
\smallskip
\item{[36]} R. D. Sorkin, ``A Finitary Substitute for Continuous Topology?'',
{\it Int. J. Th. Phys.} {\bf 30}, 923--47 (1991)
\smallskip
\item{[37]} S. M. Lewis, {\it Regge Calculus: Applications to Classical and
Quantum Gravity}, Ph.D. Thesis Univ. of Maryland at College Park (1982).
\smallskip
\item{[38]} J. W. Barrett and P. E. Parker, ``Smooth Limits of
Piecewise-Linear Approximations,'' {\it J. Approx. Theory} {\bf 76},
107--22 (1994).
\smallskip
\item{[39]} M. Ro\v cek and R. M. Williams, ``Quantum Regge Calculus,''
{\it Phys. Lett. \bf 104B}, 31--7 (1981); ``Introduction to Quantum
Regge Calculus,'' in {\it Quantum Structure of Space and Time}, eds. M. J.
Duff and C. J. Isham (Cambridge: Cambridge University Press, 1982); ``The
Quantization of Regge Calculus,'' {\it Z. Phys. \rm C \bf 21}, 371--81
(1984).
\smallskip
\item{[40]} T. Piran and A. Strominger, ``Solutions of the Regge
Equations,'' {\it Class. Quantum Grav. \bf 3}, 97--102 (1986)
\smallskip
\item{[41]} C. P. Rourke and B. J. Sanderson, {\it Introduction to
Piecewise-Linear Topology\/} (Springer, 1982), proposition 2.9

}

\vfill
\eject

\centerline{\bf Figure Captions}
\bigskip\bigskip
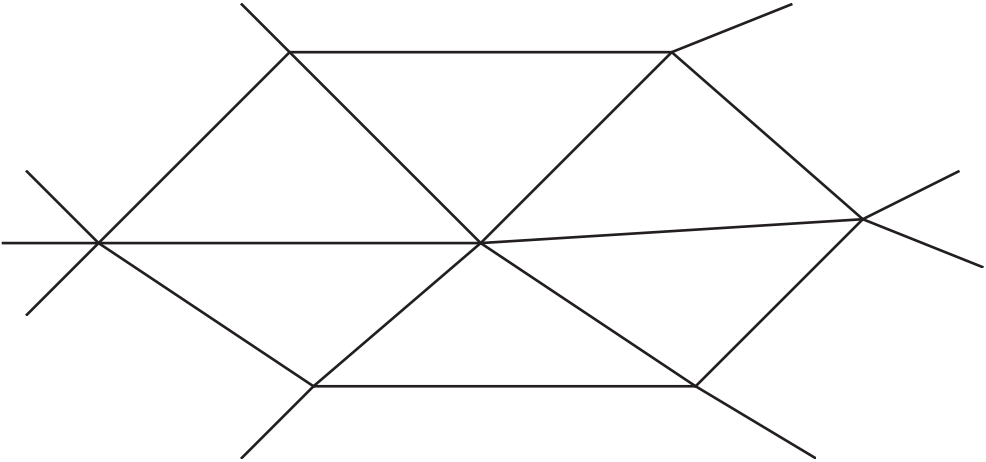
\item{\bf Fig.~1} Part of a 2-dimensional simplicial spacelike surface
surrounding a vertex.
\bigskip
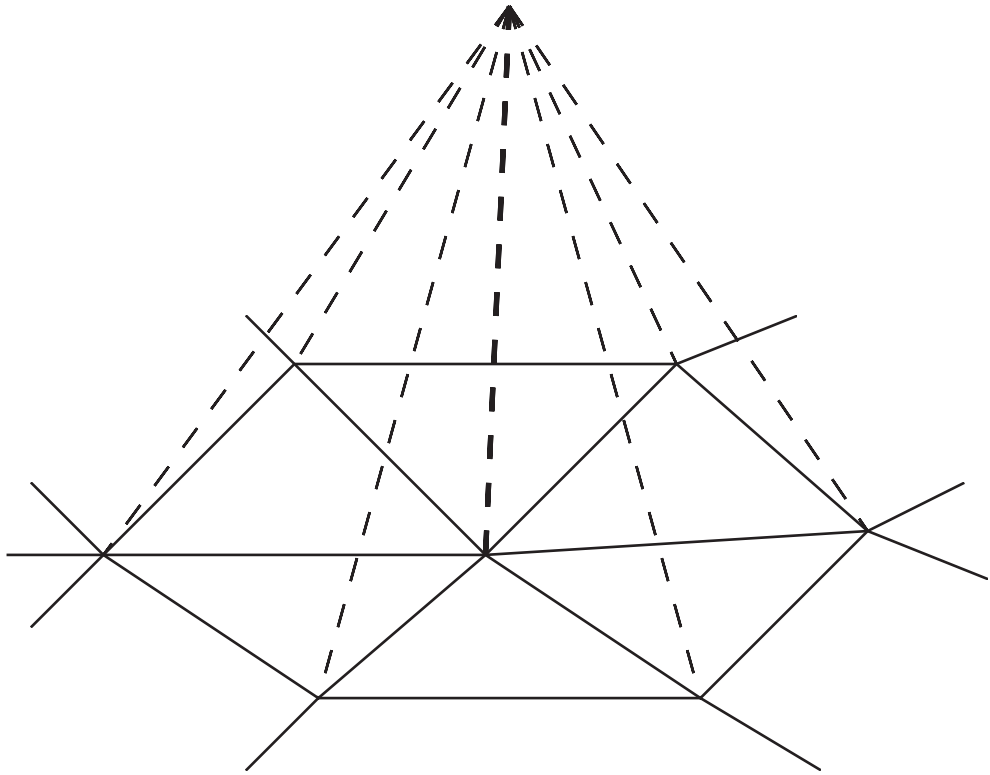
\item{\bf Fig.~2} A (2+1)--dimensional ``tent'' erected over the vertex
shown in Fig.~1. The tent consists of six tetrahedra sharing a common
``vertical'' edge (darkened dashed edge).
\bigskip
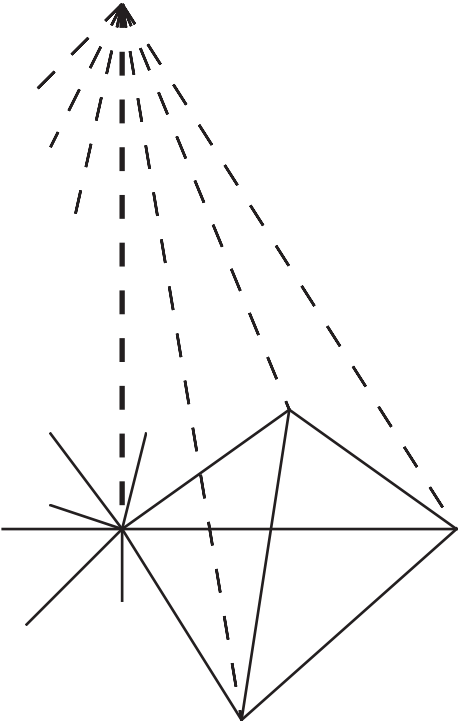
\item{\bf Fig.~3} A section of a (3+1)--dimensional ``tent'' erected over
a vertex in a spacelike hypersurface.  For each tetrahedron in the
hypersurface, a 4-simplex is constructed with apex at the new vertex. Due
to the complexity of the diagram we show here only one of the numerous
simplices sharing the common ``vertical'' edge.
\bigskip
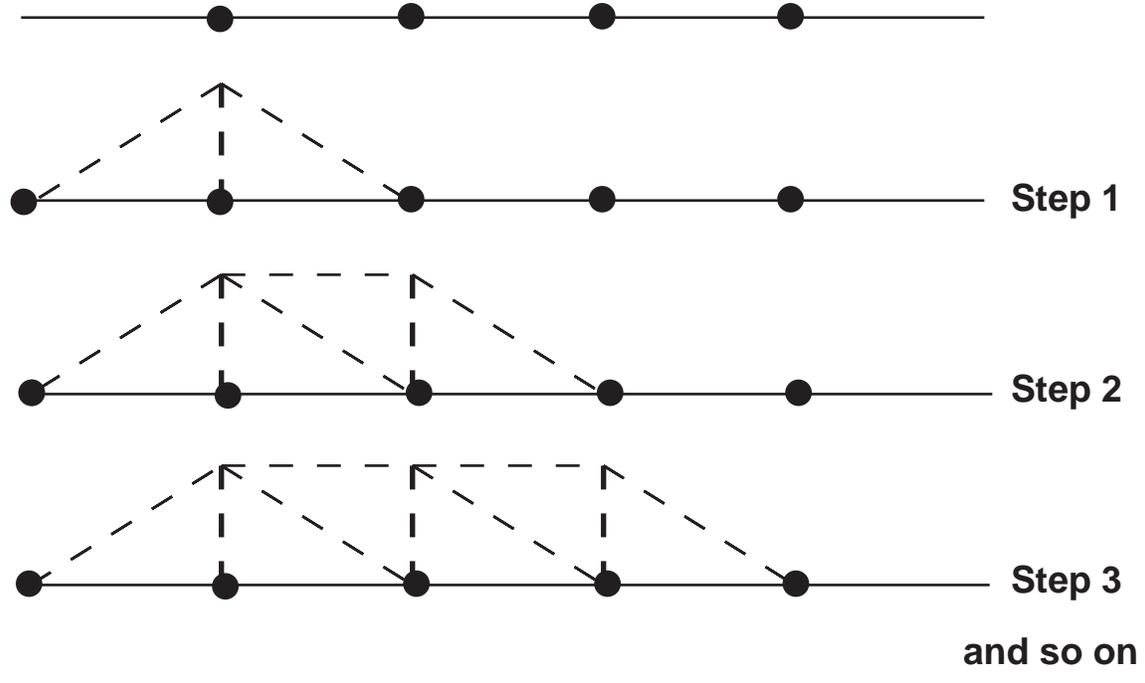
\item{\bf Fig.~4} Evolution of a surface in 1+1 dimensions, by advancing
the vertices one--by--one.
\bigskip
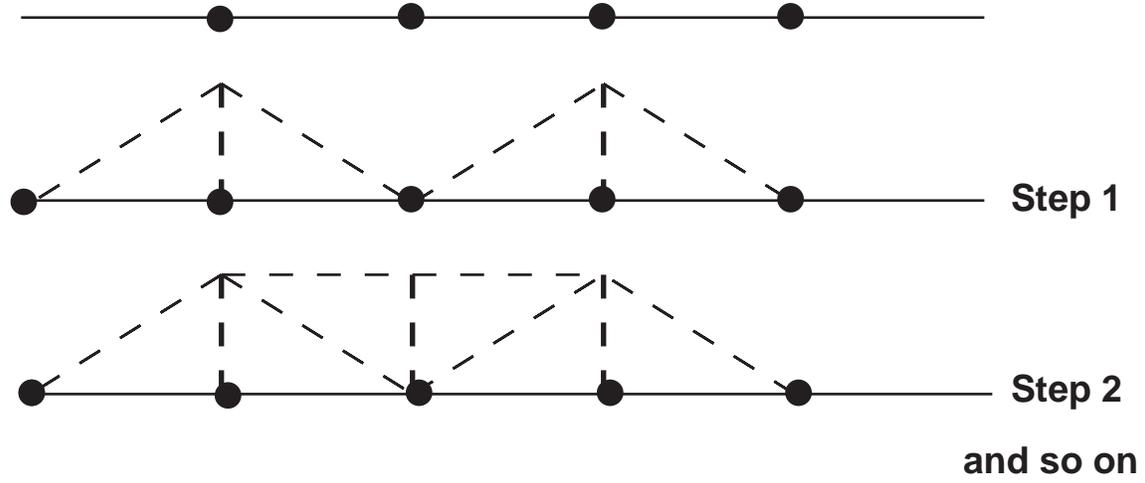
\item{\bf Fig.~5} Evolution of a surface in 1+1 dimensions, by advancing
the vertices in
parallel (cf.~Fig.~4).
\bigskip
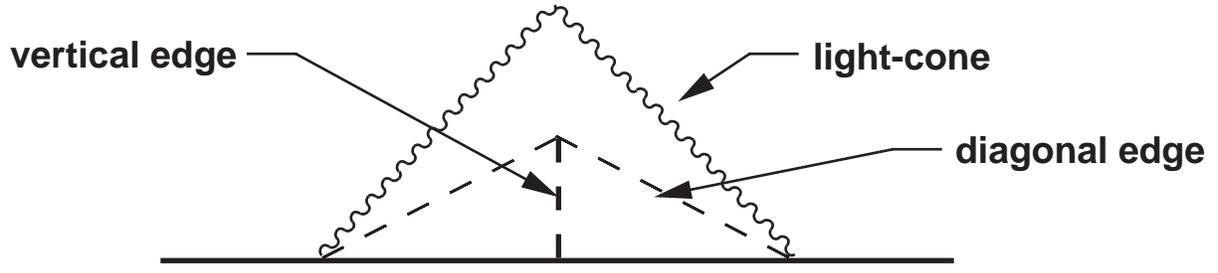
\item{\bf Fig.~6} To ensure an accurate evolution step (in accordance with
the Courant limit) we must choose the lapse and shift conditions so that
the ``tent'' lies within the ``light cone on'' its spacelike base as
illustrated here in its (1+1)--dimensional form.
(Notwithstanding the implicit character of our scheme, we believe it will
still be subject to a Courant limit as a result of its strict locality.)
\bigskip
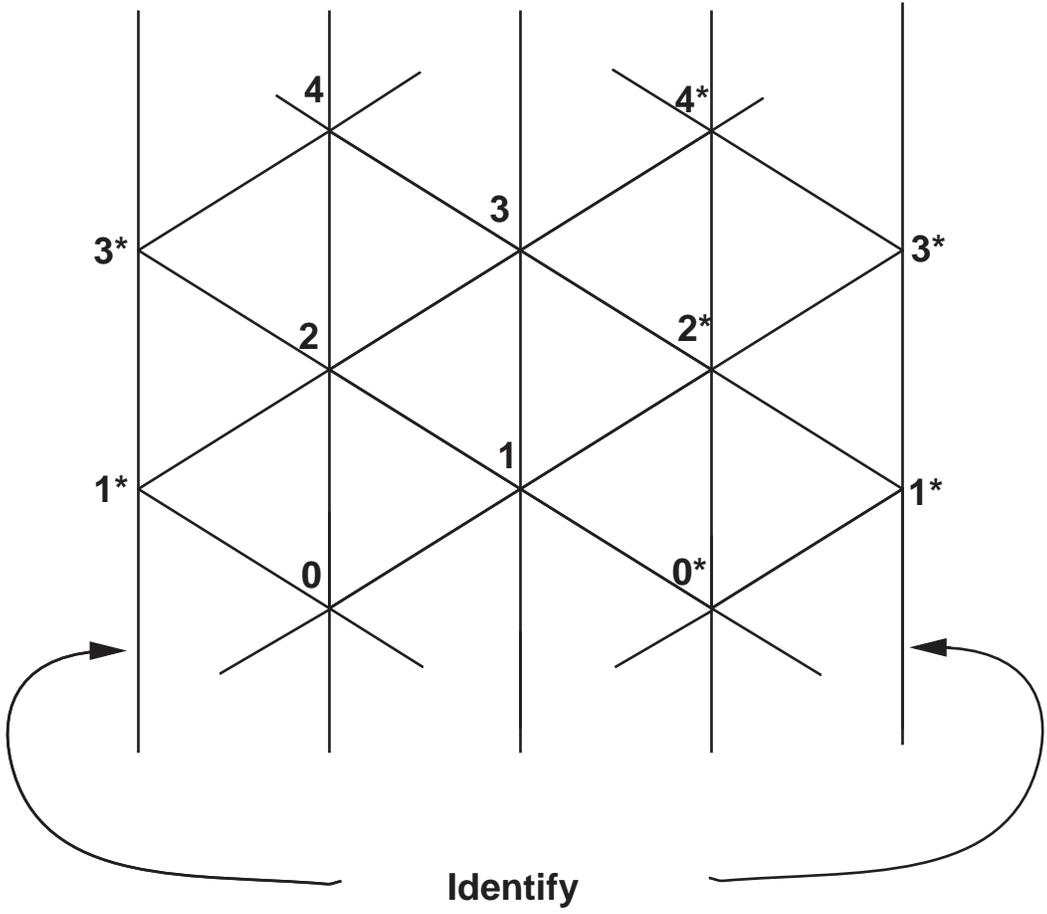
\item{\bf Fig.~7} Sorkin's lattice for $S^1\times R$.
\bigskip
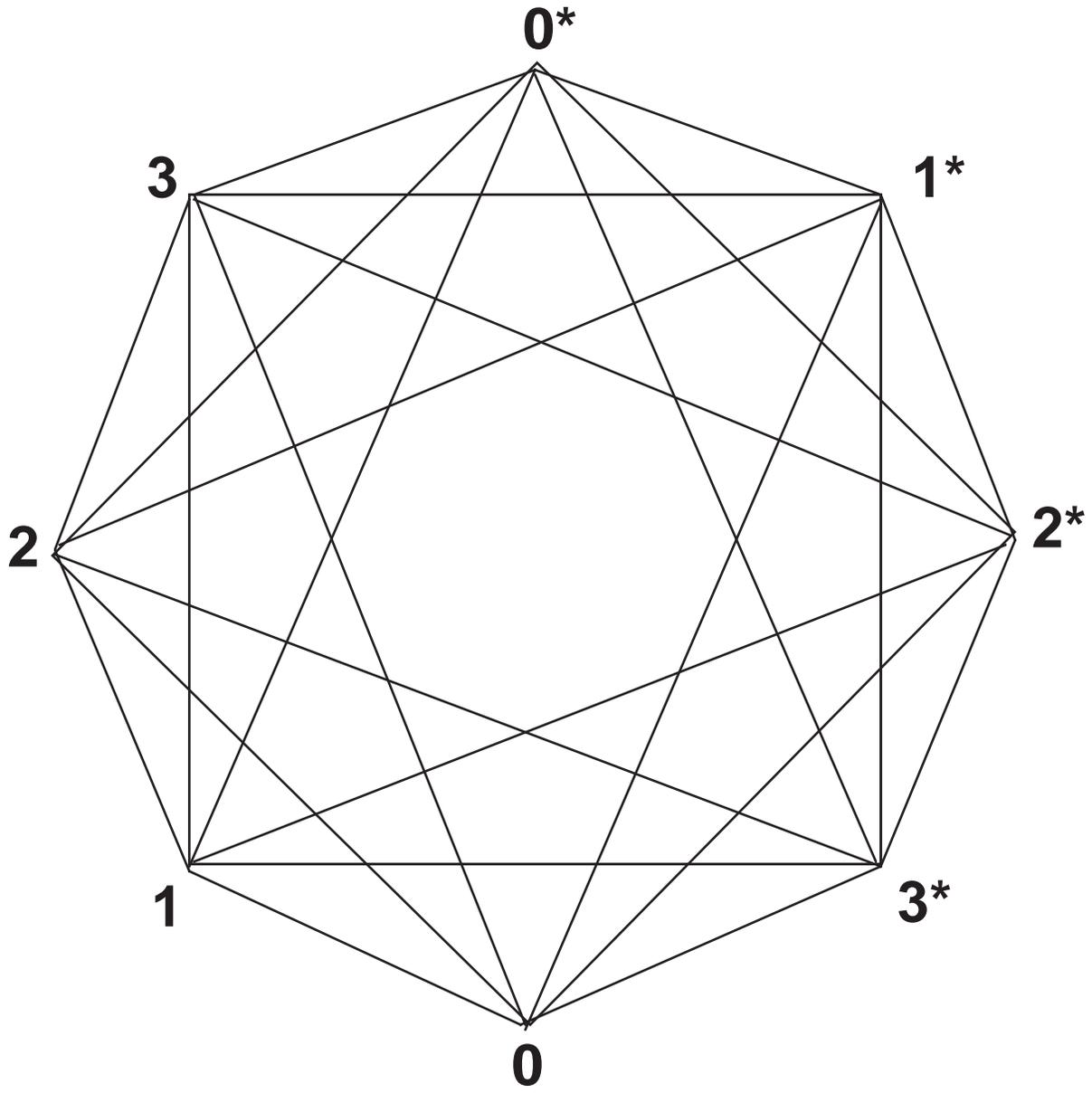
\item{\bf Fig.~8} The 16--cell tessellation $\beta_4$ of $S^3$.
\bigskip
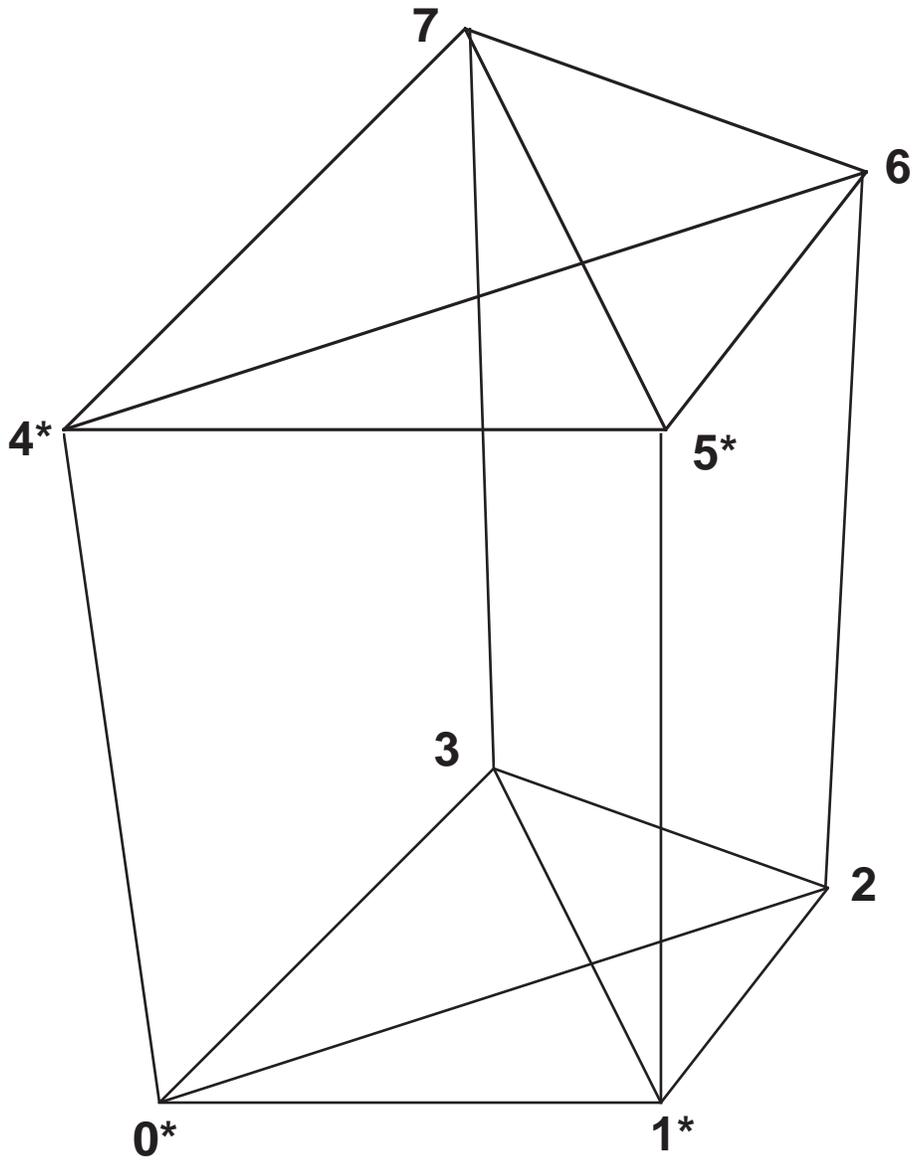
\item{\bf Fig.~9} A prism used in the evolution of $\beta _4$.  It consists
of two spacelike tetrahedra with corresponding vertices joined by timelike
edges.
\bigskip
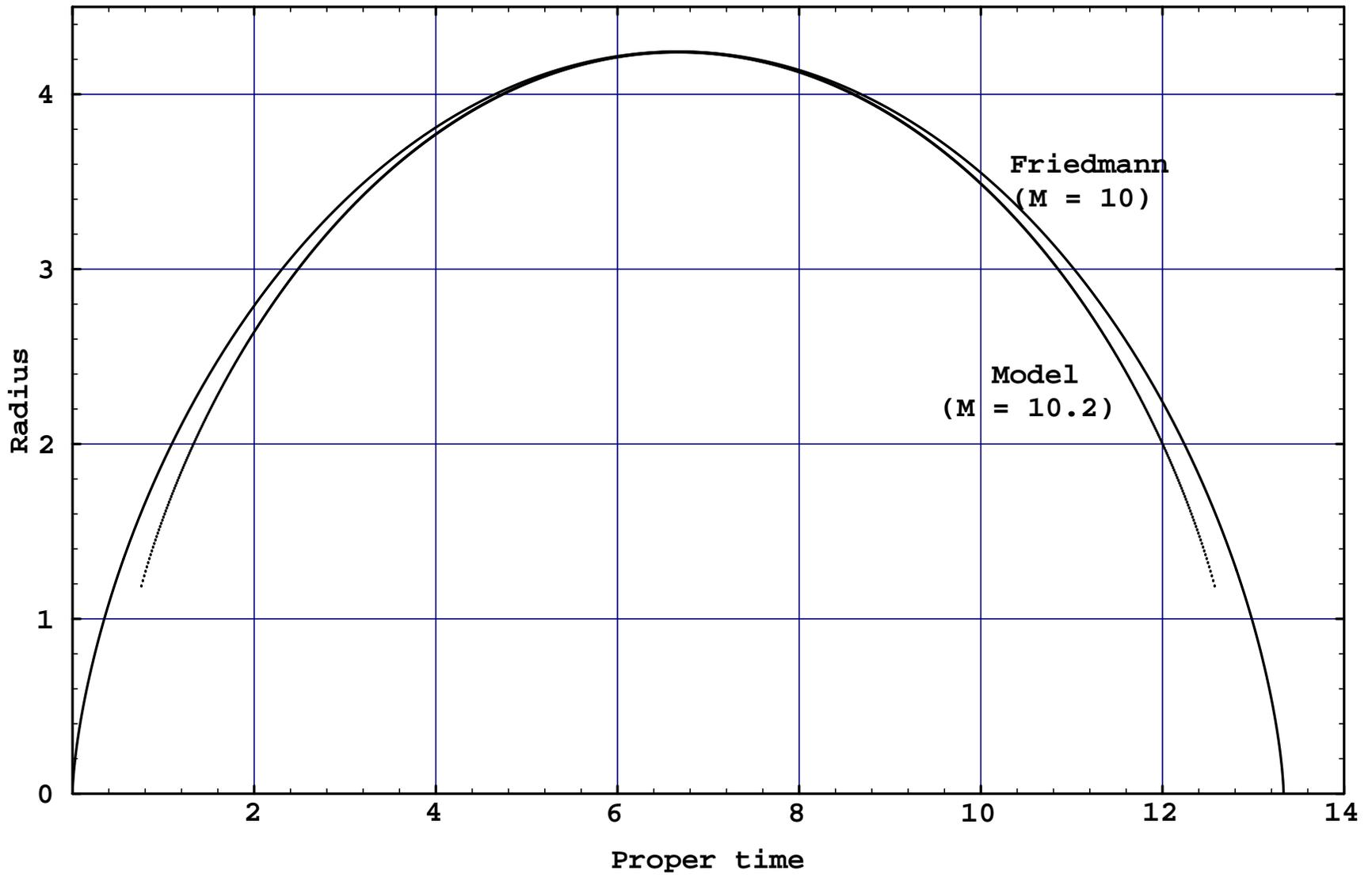
\item{\bf Fig.~10} Comparison of the effective radius $a$ (as defined in
the text, see equation 6.11) of the 600-cell model (lower, dotted line)
with the scale factor of the Friedmann universe (upper, solid line) as
functions of proper time. The Regge model has $v=.0102$ and $M=10.2$, and
$a$ is plotted versus proper time elapsed at the centre of a tetrahedron
(see text, equation 6.12); the comparison Friedmann universe has
$M=10$. (Units are such that $c = G = 1$.)

\bye